\documentclass[12pt]{iopart}
\usepackage{graphicx}
\usepackage{longtable}
\usepackage{color}
\usepackage{dcolumn}
\usepackage{multirow}
\usepackage{diagbox}
\usepackage{epstopdf}

\newcommand{\idest}[1]
{
  \textit{i.e.~}#1
}

\newcommand{\ninej}[9]
{
  \left\{ \begin{array}{ccc} #1 & #2 & #3 \\
    #4 & #5 & #6 \\ #7 & #8 & #9 \end{array} \right\}
}
\newcommand{\sixj}[6]
{
  \left\{ \begin{array}{ccc} #1 & #2 & #3 \\
    #4 & #5 & #6 \end{array} \right\}
}

\begin{document}

\title{Optical trapping of ultracold dysprosium atoms: transition probabilities, dynamic dipole polarizabilities and van der Waals $C_6$ coefficients}

\author{H. Li$^{1}$, J.-F. Wyart$^{1,2}$, O. Dulieu$^{1}$, S. Nascimb{\`e}ne$^{3}$ and M. Lepers$^{1}$}
\address{${}^{1}$Laboratoire Aim{\'e} Cotton, CNRS, Universit{\'e} Paris-Sud, ENS Cachan, Universit{\'e} Paris-Saclay, 91405 Orsay, France}
\address{${}^{2}$LERMA, Observatoire de Paris-Meudon, PSL Research University, Sorbonne Universit{\'e}s, UPMC Univ.~Paris 6, CNRS UMR8112, 92195 Meudon, France}
\address{$^{3}$Laboratoire Kastler Brossel, Coll\`ege de France, CNRS, ENS-PSL Research University, UPMC-Sorbonne Universit\'es, 11 place Marcelin Berthelot, 75005 Paris}
\ead{maxence.lepers@u-psud.fr}

\begin{abstract}
The efficiency of optical trapping of ultracold atoms depend on the atomic dynamic dipole polarizability governing the atom-field interaction. In this article, we have calculated the real and imaginary parts of the dynamic dipole polarizability of dysprosium in the ground and first excited level. Due to the high electronic angular momentum of those two states, the polarizabilities possess scalar, vector and tensor contributions that we have computed, on a wide range of trapping wavelengths, using the sum-over-state formula. Using the same formalism, we have also calculated the $C_6$ coefficients characterizing the van der Waals interaction between two dysprosium atoms in the two lowest levels. We have computed the energies of excited states and the transition probabilities appearing in the sums, using a combination of \textit{ab initio} and least-square-fitting techniques provided by the Cowan codes and extended in our group. Regarding the real part of the polarizability, for field frequencies far from atomic resonances, the vector and tensor contributions are two-order-of-magnitude smaller than the scalar contribution, whereas for the imaginary part, the vector and tensor contributions represent a noticeable fraction of the scalar contribution. This offers the possibility to control the decoherence and trap losses due to spontaneous emission.
\end{abstract}


\section{Introduction}

In the field of ultracold gases,~\idest{with} temperatures below 1 milli-kelvin, those containing particles carrying a dipole moment, so-called \textit{dipolar gases}, have attracted tremendous interest during the last years, due to their possibility of exploring strongly correlated matter, with the presence of the long-range, anisotropic dipole-dipole interaction \cite{baranov2008, lahaye2009}. In contrast with the short-range and isotropic van der Waals interaction, often approximated by contact potentials \cite{dalfovo1999}, the dipole-dipole interaction drastically modifies the properties of ultracold gases \cite{goral2000, stuhler2005, lahaye2007, bismut2012, depaz2013, baumann2014, aikawa2014b}, for example by inducing quantum-chaotic scattering between atoms \cite{frisch2014, maier2015}. Dipolar gases are also promising candidate systems for quantum information and quantum simulation \cite{lukin2001, micheli2006, bloch2012, galitski2013}.

Dipolar gases can contain different kinds of particles, whose properties can be tailored using electromagnetic fields. Firstly, electric dipole moments can be induced with external electric fields, either in highly-excited, so-called Rydberg atoms \cite{gallagher1994, anderson1998, afrousheh2004, li2005, vogt2006, brion2007, vanditzhuijzen2008, gaetan2009, saffman2010, thaicharoen2016}, or in heteronuclear alkali-metal diatomic molecules \cite{micheli2007, byrd2012b, zuchowski2013, lepers2013, vexiau2015}. Some of them have been recently produced in their lowest rovibrational and even hyperfine level, \idest{LiCs} \cite{deiglmayr2008}, KRb \cite{ni2008, ospelkaus2010b, aikawa2010}, RbCs \cite{takekoshi2014, molony2014}, NaK \cite{park2015} and NaRb \cite{zhu2016}. Open-shell polar molecules such as OH \cite{stuhl2012}, SrF \cite{barry2014}, YO \cite{yeo2015}, RbSr \cite{pasquiou2013}, which also possess a (weak) magnetic dipole moment, offer even better possibilities of control. Alternatively, ultracold gases of strong magnetic dipoles have also been produced with chromium \cite{griesmaier2005, naylor2015}, high-atomic-number (high-$Z$) lanthanides \cite{hancox2004, hemmerling2014}, including dysprosium (Dy) \cite{leefer2010, lu2010, maier2014}, erbium \cite{ban2005, mcclelland2006, frisch2012}, holmium \cite{miao2014} and thulium \cite{sukachev2010}. The formation of erbium molecules Er$_2$ have also been reported \cite{frisch2015}. Beyond the scope of dipolar gases, the specific structure of optical transitions in lanthanide atoms could be used to efficiently emulate synthetic gauge fields  \cite{cui2013}, as recently observed in Ref.~\cite{burdick2016}.

Among neutral atoms, dysprosium presents the strongest magnetic dipole moment, equal to 10 Bohr magnetons ($\mu_B$). This is due to the four unpaired $f$ electrons in the ground-level configuration [Xe]$4f^{10}6s^2$. Moreover, the excited electronic configurations which are close in energy to the lowest one \cite{brewer1983}, result in a rich energy spectrum, which is not yet completely understood \cite{NIST_ASD, nave2000, wyart2016}. Dysprosium also presents the particularity of having a pair of quasi-degenerate opposite-parity energy levels with the same electronic angular momentum $J=10$, which were used for precision measurements \cite{budker1993, budker1994, cingoz2007, leefer2013}. Because the ground and first-excited levels of dysprosium belong to the same configuration and the same LS manifold,~\idest{$^5I$,} they possess very similar electronic properties, which make them suitable candidates for optical-clock transitions \cite{kozlov2013}. Finally, along with erbium \cite{aikawa2012, aikawa2014}, dysprosium presents several bosonic and fermionic stable isotopes which allowed for the production of Bose-Einstein condensates and degenerate Fermi gases \cite{lu2011b, lu2012, tang2015}.

In this context, it is crucial to deeply understand and control how the atoms are trapped by electromagnetic fields \cite{grimm2000}. The efficiency of the trapping process is determined by the atom-field interaction, and the corresponding ac-Stark shift, which depends on the (complex) dynamic dipole polarizability of the atoms. The real part of the polarizability yields the potential energy exerted on the atomic center of mass, while the imaginary part yields the photon-scattering rate due to spontaneous emission. Because the wave functions associated with the unpaired $4f$ electrons are anisotropic, the ac-Stark shift comprises scalar, vector and tensor contributions, and so it depends on the atomic Zeeman sublevel $M_J$ and the polarization of the electromagnetic field \cite{manakov1986, lepers2014}. In this respect, the vector and tensor contributions also determine the strength of the Raman coupling between atomic Zeeman sublevels \cite{cui2013}. In literature, there exist a few theoretical values of the real part of the scalar and tensor static polarizabilities \cite{chu2007, lide2012, dzuba2014, dzuba2016}, which are in good agreement with experimental values \cite{rinkleff1994, ma2015}. By contrast the dynamic polarizability measured in a 1064-nm optical trap \cite{lu2011b} is significantly smaller then the theoretical one \cite{dzuba2011}.

In this article, we calculate the real and imaginary parts of the dynamic dipole polarizability for the ground and first excited level of dysprosium, on a wide range of frequencies of the trapping field. We give the scalar, vector and tensor contributions to the polarizability, and the useful formulas to deduce the potential energy and photon-scattering rates in the most frequently used field polarizations $\sigma^\pm$ and $\pi$. We also calculate the various $C_6$ coefficients characterizing the van der Waals dispersion interaction between two dysprosium atoms. To get all this information, we take advantage of the flexibility of the sum-over-state formula for polarizability, inherent to second-order perturbation theory. This formula is particularly well adapted to high-$Z$ lanthanide atoms, whose spectrum consists of a few strong transitions in a forest of weak transitions \cite{cowan1981}. In addition to the polarizability, we also give precious information on the spectroscopy of dysprosium, whose transition energies and transition dipole moments are computed using a combination of \textit{ab initio} and least-square fitting techniques provided by the Cowan codes \cite{cowan1981, wyart2011}. Moreover, in order to adjust experimental and theoretical transition probabilities, we employ the systematic technique that we set up in our previous works on Er$^+$ \cite{lepers2016}.

The article is outlined as follows. Section \ref{sec:eltp} presents in details the results of our electronic-structure calculations, including energy levels (subsections \ref{sub:1st-ev} and  \ref{sub:1st-odd}) and transition probabilities (subsection \ref{sub:tp}).  Then the results for the dynamic dipole polarizabilities and $C_6$ coefficients for the two lowest levels are reported in Section \ref{sec:pc}. Section \ref{sec:conclu} contains concluding remarks.

\section{Energy levels and transition probabilities\label{sec:eltp}}

Our electronic-structure calculations were carried out with the Racah-Slater method implemented in the Cowan codes \cite{cowan1981}, and which were described in our previous papers \cite{wyart2009, wyart2011, lepers2014, lepers2016}. Briefly they consist of three steps:
\begin{enumerate}
  \item Energies and transition probabilities are computed using a Hartree-Fock method including relativistic corrections and combined with configuration interaction (HFR+CI). For each parity and each value of the total electronic angular momentum $J$, the the Hamiltonian operator is a combination of angular terms, calculated using Racah algebra, and radial integrals, for example Coulombic or spin-orbit integrals. In addition, transition probabilities depend on monoelectronic transition dipole moments (MTDMs) for each pair of configurations. 
  \item The radial integrals are treated as adjustable parameters, in order to fit the theoretical energies to the experimental ones by a least-square procedure. 
  \item Similarly, the MTDMs are treated as adjustable parameters, in order to fit the theoretical transition probabilities to the experimental ones \cite{ruczkowski2014, bouazza2015}.
\end{enumerate}
For neutral dysprosium (Dy I), the experimental energies are published in the NIST database \cite{NIST_ASD}, constructed from the critical compilation of Martin \textit{et.~al.}~\cite{martin1978}, and from Ref.~\cite{nave2000} which is posterior to the compilation. For even-parity levels, we also use the yet unpublished work \cite{wyart2016}. Note that for bosonic isotopes, the nuclear spin is $I=0$, and there is no hyperfine structure. By contrast, the fermionic isotopes $^{161}$Dy and $^{163}$Dy possess a nuclear spin $I=5/2$, but the resulting hyperfine structure is not considered in the article.

\subsection{Energy levels of even parity \label{sub:1st-ev}}

The ground level of dysprosium is of even parity with the configuration $[\mathrm{Xe}]4f^{10}6s^{2}$, and total electronic angular momentum $J=8$. (In what follows, we will omit the confiration of the xenon core [Xe].) The orbital $L=6$ and spin $S=2$ angular momenta are also good quantum numbers up to 94 \%. Table \ref{tab:even} presents a comparison between our theoretical energies and Land{'e} g-factors with their experimental counterparts. All levels can be labeled in the LS coupling scheme. The levels at $E^\mathrm{exp}=13170.38$ and 15636.87 cm$^{-1}$, not present in the NIST database \cite{NIST_ASD}, come from the unpublished list of Ref.~\cite{wyart2016}. In that work, a careful modeling of the even-parity levels including $[\mathrm{Xe}]4f^{10}6s^{2}$, $[\mathrm{Xe}]4f^{10}5d6s$ and $[\mathrm{Xe}]4f^{9}6s^{2}6p$ configurations establishes that the two $^5F$ of Table \ref{tab:even} necessarily belong the $[\mathrm{Xe}]4f^{10}6s^{2}$ configuration. The agreement between theory and experiment is very good, except for the Land\'e factors of the two highest levels, which indicates that the latter are perturbed by excited configurations. The set of least-square fitted parameters used in this calculation is given in the appendix (see Table \ref{tab:parev}).

\begin{table}
	\caption{Comparison of energies $E$ through the quantity $\Delta E = E^\mathrm{exp} - E^\mathrm{th}$ and Land\'e g-factors $g_L$ of Dy I even-parity levels of the lowest electronic configuration [Xe]$4f^{10}6s^2$. The superscript {}''exp" stands for experimental values which are taken from \cite{NIST_ASD, nave2000}. The superscript {}''th" stands for the theoretical values from the present parametric calculations (see fitted parameters in Table \ref{tab:parev}). The {}``2'' in the term $^3K2$ is used to distinguish the $^3K$ terms coming from different parent terms of the $4f^9$ core (and similarly for $^3H4$) \cite{cowan-kramida}. For those two levels, the numbers between parentheses in the last column, give the total percentage of $^3K$ and $^3H$ characters respectively.}
	\label{tab:even}
	\par\centering
	\begin{tabular}{rrrrrrl}
	\hline 
	Term & $J$ & $E^\mathrm{exp}$ (cm$^{-1}$) & $\Delta E$ (cm$^{-1}$) & $g_L^\mathrm{exp}$ & $g_L^\mathrm{th}$ & \% leading term \\
	\hline                                                                   
	$^5I$ &   8 &        0        &            -13 &          1.242 &             1.243 & 94 \\
	$^5I$ &   7 &     4143.23     &             13 &          1.173 &             1.175 & 98 \\ 
	$^5I$ &   6 &     7050.61     &              7 &          1.072 &             1.073 & 96 \\
	$^5I$ &   5 &     9211.58     &            0.4 &          0.911 &             0.911 & 92 \\
	$^5I$ &   4 &    10925.25     &             -9 &          0.618 &             0.618 & 91 \\
	$^5F$ &   5 &    13170.38     &              7 &          1.358 &             1.366 & 84 \\
	$^5F$ &   4 &    15636.87     &             -6 &          1.34  &             1.339 & 93 \\
	$^3K2$ &  8 &    19019.15     &              3 &          1.14  &             1.107 & 58 (75) \\
	$^3H4$ &  6 &    24062.88     &             -2 &          1.217 &             1.176 & 41 (85) \\
	\hline 
	\end{tabular}
\end{table}

\subsection{Energy levels of odd parity \label{sub:1st-odd}}

In the odd parity, the electronic configurations included in our model are the two lowest ones $4f^{10}6s6p$ and $4f^{9}5d6s^{2}$ \cite{brewer1983}, which are connected to the ground-state configuration $4f^{10}6s^{2}$ by electric dipole (E1) transitions. Therefore, in our model, we neglect the configuration interaction with other odd-parity configurations, especially $4f^{9}5d^{2}6s$, as it results in numerous levels, making the least-square calculation hard to converge. By contrast, the first parametric study of odd-parity levels was performed with configurations with a limited number of LS terms of the $4f^n$ core, including configuration interaction with $4f^95d^26s$; but such a truncation strongly damaged the quality of the Hamiltonian eigenvectors \cite{wyart1974}. In the present study, 126 odd-parity levels were fitted to their known experimental counterparts \cite{NIST_ASD, nave2000}, using 20 free energetic parameters, giving a 44-cm$^{-1}$ standard deviation. 

A comparison between theoretical and experimental levels is displayed in Table \ref{tab:oddlev}, while the fitted parameters are given in Table \ref{tab:parod} of the appendix. All energies are given relative to the $4f^{10}6s^{2}$ $^5I_8$ ground level. Despite the absence of the $4f^95d^26s$ configuration, whose lowest classified levels is at 18472.71 cm$^{-1}$, the agreement is very satisfactory. On the contrary, a poor agreement, especially on Land\'e factors, reflects local perturbations by the $4f^95d^26s$ configuration.

\begin{longtable}{rrrlllr}
	\caption{Same as Table \ref{tab:even} for Dy I odd-parity levels. The theoretical values $E^\mathrm{th}$, the Land\'e g-factors	$g_L^\mathrm{th}$ and the percentage of configurations and LS terms are derived by means of the RCG code with the parameter set reported in Table \ref{tab:parod}. In the configuration notations, $A$ stands for $4f^{10}$, $B$ for $4f^{9}$, $ds^2$ for $5d6s^2$, $sp$ for $6s6p$. The lower-case letters or Arabic numbers appearing in the seventh column correspond to different possible parent terms \cite{cowan1981}. The terms in parentheses are associated with the core configurations $A$ or $B$.}
	\label{tab:oddlev}                             
	\endfirsthead
	\caption{Odd parity levels of Dy I (continued)} \\
	\hline
	$E^\mathrm{exp}$ & $E^\mathrm{th}$ & $\Delta E$ & \multirow{2}{*}{$g_L^\mathrm{exp}$} & \multirow{2}{*}{$g_L^\mathrm{th}$} & Leading & \% leading   \\
	(cm$^{-1}$) & (cm$^{-1}$) & (cm$^{-1}$) & & & configuration & LS term \\
	\hline
	\endhead
	\hline
	$E^\mathrm{exp}$ & $E^\mathrm{th}$ & $\Delta E$ & \multirow{2}{*}{$g_L^\mathrm{exp}$} & \multirow{2}{*}{$g_L^\mathrm{th}$} & Leading & \% leading   \\
	(cm$^{-1}$) & (cm$^{-1}$) & (cm$^{-1}$) & & & configuration & LS term \\
	\hline
	&          &             & $J=2$ &       &       &            \\
	28407.01   &  28407.9    &  -1   & 0.06  & 0.029 & $A-sp$    &   91 $A-sp   \, (^5I) ^7H$   \\ 
	\hline 
	&    & & $J=3$ & & &     \\
	15254.94   &  15285.5    & -31   & 0.77  & 0.777 & $B-ds^2$  &   73 $B-ds^2 \, (^6H) ^7H$   \\ 
	23824.68   &  23753.0    &  72	 & 0.68  & 0.665 & $B-ds^2$  &   61 $B-ds^2 \, (^6H) ^5H$   \\   
	24668.59   &  24642.8    &  26	 & 1.29  & 1.227 & $B-ds^2$  &   22 $B-ds^2 \, (^6H) ^5F$  \\    
	26607.16   &  26647.2    & -40   & 0.58  & 0.464 & $A-sp$    &   55 $A-sp   \, (^5I) ^7I$   \\     
	26886.01   &  26952.9    & -67   & 1.02  & 1.075 & $B-ds^2$  &   33 $B-ds^2 \, (^6F) ^5G$   \\    
	27321.26   &  27350.0    & -29   & 0.58  & 0.551 & $A-sp$    &   31 $A-sp   \, (^5I) ^7H$   \\    
	27643.57   &  27617.4    &  26	 & 1.17  & 0.726 & $B-ds^2$  &   61 $B-ds^2 \, (^6F) ^5H$   \\    
	28694.51   &  28720.7    & -26   & 0.55  & 0.551 & $A-sp$    &   27 $A-sp   \, (^5I) ^5Ha$   \\
	\hline
	&  & & $J=4$ & & &     \\  
	13952.00   &  13984.0    & -32   & 1.082 & 1.072 & $B-ds^2$  &   67 $B-ds^2 \, (^6H) ^7H$   \\ 
	16069.98   &  15986.5    &  83   & 1.62  & 1.627 & $B-ds^2$  &   46 $B-ds^2 \, (^6F) ^7P$   \\   
	16412.80   &  16348.1    &  65	 & 1.51  & 1.500 & $B-ds^2$  &   32 $B-ds^2 \, (^6H) ^7F$   \\    
	20430.11   &  20423.4    &   7	 & 1.28  & 1.262 & $B-ds^2$  &   36 $B-ds^2 \, (^6H) ^5G$   \\     
	20474.99   &  20457.6    &  17	 & 1.30  & 1.373 & $B-ds^2$  &   30 $B-ds^2 \, (^6H) ^7F$   \\    
	22099.06   &  22058.3    &  41	 & 1.059 & 1.074 & $B-ds^2$  &   36 $B-ds^2 \, (^6H) ^5H$   \\    
	22938.03   &  22925.4    &  13	 & 1.07  & 1.063 & $B-ds^2$  &   25 $B-ds^2 \, (^6H) ^5F$   \\    
	23686.81   &  23663.8    &  23	 & 0.767 & 0.743 & $B-ds^2$  &   50 $B-ds^2 \, (^6H) ^5I$   \\   
	24841.04   &  24897.4    & -56   & 0.90  & 0.881 & $A-sp$    &   39 $A-sp   \, (^5I) ^7H$   \\
	25203.92   &  25272.1    & -68   & 1.242 & 1.245 & $B-ds^2$  &   32 $B-ds^2 \, (^6F) ^5G$   \\
	25687.20   &  25726.2    & -39   & 0.94  & 0.829 & $A-sp$    &   24 $A-sp   \, (^5I) ^7I$   \\
	26440.41   &  26446.4    &  -6	 & 1.046 & 1.040 & $B-ds^2$  &   55 $B-ds^2 \, (^6F) ^5H$   \\
	26662.41   &  26716.0    & -54   & 0.59  & 0.491 & $A-sp$    &   66 $A-sp   \, (^5I) ^7K$   \\
	26998.27   &  27018.3    & -20   & 0.86  & 0.865 & $A-sp$    &   41 $A-sp   \, (^5I) ^7H$   \\ 
	27659.02   &  27609.3    &  50	 & 1.17  & 1.171 & $B-ds^2$  &   34 $B-ds^2 \, (^6F) ^5G$   \\
	27751.46   &  27746.0    &   5	 & 0.81  & 0.742 & $A-sp$    &   45 $A-sp   \, (^5I) ^3H$   \\
	28923.05   &  28966.4    & -43   & 0.78  & 0.778 & $A-sp$    &   25 $A-sp   \, (^5I) ^3H$   \\
	33324.06   &  33281.2    &  43	 & 0.89  & 0.890 & $A-sp$    &   39 $A-sp   \, (^5I) ^5Hb$  \\
	33952.33   &  34025.8    & -73   & 1.30  & 1.318 & $B-ds^2$  &    9 $B-ds^2 \, (^4G) ^5D4$  \\
	34038.46   &  34007.8    &  31	 & 1.30  & 1.305 & $A-sp$    &   22 $A-sp   \, (^5F) ^3G$   \\
	34486.89   &  34487.9    &  -1	 & 1.213 & 1.344 & $A-sp$    &   21 $A-sp   \, (^5F) ^5Fa$   \\
	34720.68   &  34704.0    &  17	 & 0.761 & 0.659 & $A-sp$    &   43 $A-sp   \, (^5I) ^5Ib$   \\
	\hline
	& & & $J=5$ & & &     \\
	12298.55   &  12334.8    & -36   & 1.24  & 1.233 & $B-ds^2$  &   58 $B-ds^2 \, (^6H) ^7H$   \\
	14153.49   &  14131.0    &  22	 & 1.42  & 1.419 & $B-ds^2$  &   51 $B-ds^2 \, (^6H) ^7F$   \\
	16684.73   &  16664.2    &  21	 & 1.082 & 1.067 & $B-ds^2$  &   66 $B-ds^2 \, (^6H) ^7I$   \\
	17502.89   &  17506.3    &  -3	 & 1.45  & 1.426 & $B-ds^2$  &   33 $B-ds^2 \, (^6F) ^7D$   \\
	17804.24   &  17834.6    &  -30  & 1.322 & 1.320 & $B-ds^2$  &   44 $B-ds^2 \, (^6H) ^5F$   \\
	19480.87   &  19563.7    &  -83  & 1.35  & 1.334 & $B-ds^2$  &   29 $B-ds^2 \, (^6F) ^7G$   \\
	19813.98   &  19794.5    &  19	 & 1.27  & 1.281 & $B-ds^2$  &   22 $B-ds^2 \, (^6H) ^5H$   \\
	20921.55   &  20901.2    &  20	 & 1.30  & 1.121 & $B-ds^2$  &   32 $B-ds^2 \, (^6H) ^5H$   \\
	22294.88   &  22296.6    &  -2	 & 1.02  & 0.990 & $B-ds^2$  &   35 $B-ds^2 \, (^6H) ^5I$   \\
	20921.55   &  20901.2    &  20	 & 1.30  & 1.121 & $B-ds^2$  &   32 $B-ds^2 \, (^6H) ^5H$   \\
	22294.88   &  22296.6    &  -2	 & 1.02  & 0.990 & $B-ds^2$  &   35 $B-ds^2 \, (^6H) ^5I$   \\
	22524.21   &  22498.8    &  25	 & 1.04  & 1.052 & $A-sp$    &   19 $A-sp   \, (^5I) ^7H$   \\
	23552.65   &  23553.9    &  -1	 & 1.07  & 1.051 & $A-sp$    &   27 $A-sp   \, (^5I) ^7I$   \\
	24634.07   &  24637.6    &  -3	 & 1.21  & 1.214 & $B-ds^2$  &   37 $B-ds^2 \, (^6F) ^5H$   \\
	24881.87   &  24938.6    &  -57  & 0.72  & 0.753 & $B-ds^2$  &   70 $B-ds^2 \, (^6H) ^5K$   \\
	25082.02   &  24997.7    &  84	 & 1.064 & 0.992 & $A-sp$    &   41 $A-sp   \, (^5I) ^7H$   \\
	25127.52   &  25152.4    &  -25  & 1.04  & 0.843 & $A-sp$    &   42 $A-sp   \, (^5I) ^7K$   \\
	25912.63   &  25892.0    &  20	 & 0.98  & 0.979 & $A-sp$    &   30 $A-sp   \, (^5I) ^3H$  \\
	26135.21   &  26106.5    &  29	 & 1.22  & 1.214 & $B-ds^2$  &   30 $B-ds^2 \, (^6F) ^5G$  \\
	27109.93   &  27140.3    &  -30  & 1.01  & 0.993 & $A-sp$    &   23 $A-sp   \, (^5I) ^7I$   \\
	27685.87   &  27695.2    &  -9	 & 0.77  & 0.765 & $A-sp$    &   31 $A-sp   \, (^5I) ^5Ka$   \\
	29054.36   &  29112.4    &  -58  & 0.84  & 0.871 & $A-sp$    &   32 $A-sp   \, (^5I) ^3I$   \\
	30904.89   &  30885.3    &  19	 & 1.286 & 1.159 & $A-sp$    &   29 $A-sp   \, (^5I) ^5Hb$  \\
	31763.85   &  31762.6    &  1	 & 1.32  & 1.342 & $A-sp$    &   28 $A-sp   \, (^5F) ^3G$   \\
	33025.64   &  32950.1    &  75	 & 1.01  & 0.923 & $A-sp$    &   38 $A-sp   \, (^5I) ^5Ib$   \\
	33652.23   &  33639.7    &  13	 & 1.16  & 1.254 & $B-ds^2$  &   12 $B-ds^2 \, (^4G) ^5G4$   \\
	34470.70   &  34513.4    &  -43  & 0.915 & 0.715 & $A-sp$    &   48 $A-sp   \, (^5I) ^5Kb$   \\           
	\hline      
	& & & $J=6$ & & &     \\ 
	10088.80   &  10146.1     &  -57  & 1.36  & 1.357 & $B-ds^2$  &   36 $B-ds^2 \, (^6H) ^7H$   \\ 
	11673.49   &  11649.2     &  24	 & 1.392 & 1.395 & $B-ds^2$  &   49 $B-ds^2 \, (^6H) ^7F$   \\
	14970.70   &  15006.3     &  -36  & 1.24  & 1.238 & $B-ds^2$  &   42 $B-ds^2 \, (^6H) ^7I$   \\
	15862.64   &  15862.7     &  0	 & 1.257 & 1.260 & $B-ds^2$  &   51 $B-ds^2 \, (^6H) ^5G$   \\
	16591.38   &  16522.4     &  69	 & 1.348 & 1.356 & $B-ds^2$  &   59 $B-ds^2 \, (^6H) ^7G$   \\
	18172.87   &  18254.0     &  -81  & 1.34  & 1.305 & $B-ds^2$  &   22 $B-ds^2 \, (^6H) ^5H$   \\
	18561.20   &  18629.5     &  -68  & 1.27  & 1.301 & $B-ds^2$  &   20 $B-ds^2 \, (^6H) ^5H$   \\
	18711.93   &  18724.8     &  -13  & 1.172 & 1.171 & $A-sp$	 &   42 $A-sp   \, (^5I) ^3H$   \\
	19182.66   &  19157.5     &  25   & 1.036 & 1.032 & $B-ds^2$  &   63 $B-ds^2 \, (^6H) ^7K$   \\
	19856.88   &  19862.5     &  -6   & 1.35  & 1.314 & $B-ds^2$  &   38 $B-ds^2 \, (^6F) ^7H$   \\
	20554.73   &  20531.2     &  23   & 1.11  & 1.107 & $B-ds^2$  &   29 $B-ds^2 \, (^6H) ^5I$   \\
	20817.61   &  20798.4     &  19	 & 1.13  & 1.135 & $A-sp$	 &   17 $A-sp   \, (^5I) ^7I$   \\
	22286.87   &  22281.7     &  5	 & 1.15  & 1.151 & $A-sp$    &   38 $A-sp   \, (^5I) ^7H$   \\
	22633.23   &  22669.9     &  -36  & 1.29  & 1.297 & $B-ds^2$  &   54 $B-ds^2 \, (^6F) ^5G$   \\
	22956.84   &  22985.1     &  -28  & 1.06  & 1.083 & $A-sp$    &   22 $A-sp   \, (^5I) ^7K$   \\
	23464.02   &  23492.1     &  -28  & 0.96  & 0.946 & $B-ds^2$  &   67 $B-ds^2 \, (^6H) ^5K$   \\
	23687.87   &  23640.9     &  47	 & 1.076 & 1.064 & $A-sp$	 &   16 $A-sp   \, (^5I) ^7K$   \\
	24040.59   &  24026.1     &  14	 & 1.26  & 1.263 & $B-ds^2$  &   31 $B-ds^2 \, (^6F) ^5H$   \\
	24931.63   &  24936.8     &  -5	 & 1.128 & 1.115 & $A-sp$	 &   31 $A-sp   \, (^5I) ^7I$   \\
	25825.83   &  25836.0     &  -10  & 1.00  & 0.995 & $A-sp$	 &   29 $A-sp   \, (^5I) ^7K$   \\
	27199.20   &  27226.9     &  -27  & 1.16  & 1.032 & $A-sp$	 &   16 $A-sp   \, (^5I) ^3K$   \\
	28119.93   &  28065.6     &  54	 & 1.198 & 1.193 & $A-sp$	 &   46 $A-sp   \, (^5I) ^5Hb$   \\
	29447.11   &  29428.2     &  19   & 0.90  & 0.898 & $A-sp$	 &   56 $A-sp   \, (^5I) ^3K$   \\
	30778.96   &  30732.5     &  46	 & 1.17  & 1.073 & $A-sp$	 &   42 $A-sp   \, (^5I) ^5Ib$   \\
	32126.16   &  32096.3     &  30	 & 1.23  & 1.209 & $B-ds^2$  &   16 $B-ds^2 \, (^4F) ^5G3$   \\
	\hline
	& & & $J=7$ & & &    \\
	8519.21    &   8595.5    &  -76   & 1.336 & 1.338 & $B-ds^2$  &   59 $B-ds^2 \, (^6H) ^7H$    \\
	12655.13   &  12695.4    &  -40   & 1.36  & 1.356 & $B-ds^2$  &   61 $B-ds^2 \, (^6H) ^7G$    \\
	14367.81   &  14291.7    &   76   & 1.27  & 1.269 & $B-ds^2$  &   46 $B-ds^2 \, (^6H) ^7I$    \\
	15194.83   &  15243.3    &  -48   & 1.26  & 1.263 & $B-ds^2$  &   60 $B-ds^2 \, (^6H) ^5H$    \\
	16693.87   &  16659.2    &   34   & 1.22  & 1.227 & $A-sp$	 &   23 $A-sp   \, (^5I) ^5Ha$   \\
	17687.90   &  17681.1    &    6   & 1.16  & 1.152 & $B-ds^2$  &   44 $B-ds^2 \, (^6H) ^7K$    \\
	18339.80   &  18349.3    &  -10   & 1.21  & 1.197 & $B-ds^2$  &   15 $B-ds^2 \, (^6H) ^7K$    \\
	18433.76   &  18429.4    &    4   & 1.20  & 1.195 & $A-sp$	 &   23 $A-sp   \, (^5I) ^3I$    \\
	18857.04   &  18807.7    &   49   & 1.335 & 1.323 & $B-ds^2$  &   35 $B-ds^2 \, (^6F) ^7H$    \\
	19907.51   &  19904.9    &    3   & 1.23  & 1.237 & $A-sp$	 &   26 $A-sp   \, (^5I) ^7I$    \\
	20485.40   &  20501.7    &  -16   & 1.375 & 1.381 & $B-ds^2$  &   64 $B-ds^2 \, (^6F) ^7G$    \\
	20766.29   &  20720.7    &   46   & 1.16  & 1.140 & $A-sp$	 &   19 $A-sp   \, (^5I) ^5Ka$   \\
	21675.28   &  21698.5    &  -23   & 1.22  & 1.265 & $B-ds^2$  &   41 $B-ds^2 \, (^6F) ^5H$    \\
	21783.41   &  21766.6    &   17   & 1.15  & 1.110 & $B-ds^2$  &   58 $B-ds^2 \, (^6H) ^5K$    \\
	22061.29   &  22052.6    &    9   & 1.18  & 1.184 & $A-sp$	 &   28 $A-sp   \, (^5I) ^7I$    \\
	23479.77   &  23482.0    &   -2   & 1.13  & 1.135 & $A-sp$	 &   33 $A-sp   \, (^5I) ^7K$    \\
	24708.97   &  24772.3    &  -63   & 1.26  & 1.265 & $A-sp$	 &   38 $A-sp   \, (^5I) ^5Hb$   \\
	24906.86   &  24912.5    &   -6   & 1.14  & 1.147 & $A-sp$	 &   26 $A-sp   \, (^5I) ^7I$    \\
	27427.08   &  27410.6    &   17   & 1.06  & 1.067 & $A-sp$	 &   28 $A-sp   \, (^5I) ^3K$    \\
	27834.93   &  27850.4    &  -16   & 1.22  & 1.169 & $A-sp$	 &   47 $A-sp   \, (^5I) ^5Ib$   \\
	30711.72   &  30760.7    &  -49   & 1.09  & 1.068 & $A-sp$	 &   46 $A-sp   \, (^5I) ^5Kb$   \\
	31698.32   &  31700.2    &   -2   & 1.131 & 1.125 & $B-ds^2$  &   17 $B-ds^2 \, (^4I) ^5I3$   \\ 
	\hline
	& & & $J=8$ & & &    \\
	7565.61   &   7586.7    &  -21   & 1.352 & 1.356 & $B-ds^2$  &   77 $B-ds^2 \, (^6H) ^7H$    \\
	12007.12   &  11949.8    &   57   & 1.28  & 1.278 & $B-ds^2$  &   49 $B-ds^2 \, (^6H) ^7I$    \\
	14625.64   &  14683.1    &  -57   & 1.25  & 1.252 & $B-ds^2$  &   61 $B-ds^2 \, (^6H) ^5I$    \\
	15567.38   &  15556.1    &   11   & 1.31  & 1.322 & $A-sp$    &   58 $A-sp   \, (^5I) ^7H$    \\
	16288.73   &  16220.7    &   68   & 1.19  & 1.187 & $B-ds^2$  &   47 $B-ds^2 \, (^6H) ^7K$    \\
	16733.20   &  16677.7    &   55   & 1.20  & 1.198 & $A-sp$    &   33 $A-sp   \, (^5I) ^3K$    \\
	18021.89   &  18025.3    &   -3   & 1.23  & 1.230 & $A-sp$    &   23 $A-sp   \, (^5I) ^3K$    \\
	19092.30   &  19026.8    &   65   & 1.33  & 1.342 & $B-ds^2$  &   77 $B-ds^2 \, (^6F) ^7H$    \\
	19688.60   &  19655.1    &   33   & 1.22  & 1.200 & $B-ds^2$  &   47 $B-ds^2 \, (^6H) ^5K$    \\
	20341.32   &  20331.9    &   10   & 1.23  & 1.230 & $A-sp$    &   30 $A-sp   \, (^5I) ^7K$    \\
	21899.20   &  21874.7    &   25   & 1.20  & 1.210 & $A-sp$    &   33 $A-sp   \, (^5I) ^7I$    \\
	23877.74   &  23844.6    &   33   & 1.29  & 1.239 & $A-sp$    &   48 $A-sp   \, (^5I) ^5Ib$   \\
	24999.58   &  24976.9    &   23   & 1.19  & 1.172 & $A-sp$    &   36 $A-sp   \, (^5I) ^7K$    \\
	27818.00   &  27871.1    &  -53   & 1.21  & 1.154 & $A-sp$    &   45 $A-sp   \, (^5I) ^5Kb$   \\	 
	\hline
	& & & $J=9$ & & &    \\
	9990.97   &   9991.2    &    0	  & 1.32  & 1.320 & $B-ds^2$  &   86 $B-ds^2 \, (^6H) ^7I$    \\
	13495.93   &  13463.1    &   32	  & 1.23  & 1.233 & $B-ds^2$  &   60 $B-ds^2 \, (^6H) ^5K$    \\
	15972.35   &  15972.8    &    0	  & 1.29  & 1.294 & $A-sp$    &   61 $A-sp   \, (^5I) ^7I$    \\
	16717.79   &  16749.8    &  -32   & 1.24  & 1.242 & $B-ds^2$  &   62 $B-ds^2 \, (^6H) ^7K$    \\
	17727.15   &  17699.8    &   27	  & 1.25  & 1.258 & $A-sp$    &   31 $A-sp   \, (^5I) ^7I$    \\
	21838.55   &  21798.5    &   40	  & 1.25  & 1.244 & $A-sp$    &   64 $A-sp   \, (^5I) ^7K$    \\
	23736.61   &  23788.7    &  -52   & 1.22  & 1.217 & $A-sp$    &   48 $A-sp   \, (^5I) ^5Ka$   \\	 
	\hline
	& & & $J=10$ & & &    \\
	12892.76   &  12992.5    &  -99   & 1.29  & 1.294 & $B-ds^2$  &   94 $B-ds^2 \, (^6H) ^7K$    \\
	17513.33   &  17465.2    &   48   & 1.30  & 1.295 & $A-sp$	  &   94 $A-sp   \, (^5I) ^7K$    \\   
	\hline
\end{longtable}

\subsection{Transition probabilities \label{sub:tp}}

Since they depend on transition dipole moments, the transition probabilities turn out to be an efficient test for the quality of our computed eigenvectors. After the last CI calculation by RCG, the eigenvector of the level $i$ can be written $|i\rangle=\sum_p c_{ip}|p\rangle$, where $|p\rangle$ represents an electronic configuration. Then the theoretical Einstein coefficients $A_{ij}^\mathrm{th}$ characterizing the probability of spontaneous emission from level $i$ to level $j$ can be expanded
\begin{equation}
	A_{ij}^\mathrm{th} = \left( \sum_{pq} a_{ij,pq}
	\left\langle n\ell,p\right|
	\hat{r}\left|n'\ell',q\right\rangle 
	\right)^2 ,
	\label{eq:aik}
\end{equation}
in which the MTDMs $\left\langle n\ell,p\right|	\hat{r}\left|n'\ell',q\right\rangle$ are common parameters to all transitions. The configurations included in our model give rise to two possible $\hat{r}$-matrix elements: one for $(n\ell$-$n'\ell') = (6s$-$6p)$ transitions, namely $(p,q)=(4f^{10}6s^2,4f^{10}6s6p)$, and the other one for $(n\ell$-$n'\ell') = (4f$-$5d)$ transitions, namely $(p,q)=(4f^{10}6s^2,4f^{9}5d6s^2)$.

Similarly to section \ref{sub:1st-odd}, our theoretical $A$ coefficients depend on a restricted number of scaling factors (SFs) $f_m$, which are also adjusted by fitting to available experimental data \cite{wickliffe2000}. The $f_m$ can be defined from MTDMs and their computed HFR values, $f_m = \left\langle n\ell,p\right| \hat{r}\left|n'\ell',q\right\rangle / \left\langle n\ell,p\right| \hat{r}\left| n'\ell',q\right\rangle_\mathrm{HFR}$. We specify $f_1$ and $f_2$ for $(6s$-$6p)$ and $(4f$-$5d)$ transitions respectively. From Ref.~\cite{wickliffe2000}, we can get 80 transitions towards the ground level $^5I_8$ and first excited one $^5I_7$. As energy increases, especially above 30000 cm$^{-1}$, it is hard to describe accurately the energy levels only with the $4f^{10}6s6p$ and $4f^{9}5d6s^2$ configurations (see Table \ref{tab:oddlev}). Therefore we exclude from the fitting procedure these transitions with upper levels above 30000 cm$^{-1}$.

\begin{longtable}{rrrrrrl}
	\caption{Transitions excluded from the least-square fitting procedure. The letters $i$ and $j$ correspond to upper and lower levels, respectively. The superscript {}``exp" stands for experimental values
		which are taken from \cite{wickliffe2000}. The superscript {}``th" 
		stands for the theoretical values from the present parametric 
		calculations. The transition wave number $\sigma_{ij}$ is in the vacuum. The notation ($n$) stands for $\times 10^n$. A blank in the column {}``removal reason'' means that the upper level belongs neither to the $4f^{10}6s6p$ nor to the $4f^95d6s^2$ configuration.}
	\label{tab:rml}                             
	\endfirsthead
	\caption{excluded transitions (continued)} \\
	\hline
	$E_i^\mathrm{exp}$ (cm$^{-1}$) & $J_i$ & $E_j^\mathrm{exp}$ (cm$^{-1}$) & $J_j$ & $\sigma_{ij}^\mathrm{exp}$ (cm$^{-1}$) & $A_{ij}^\mathrm{exp}$ (s$^{-1}$) & removal reason \\
	\hline
	\endhead
	\hline
	$E_i^\mathrm{exp}$ (cm$^{-1}$) & $J_i$ & $E_j^\mathrm{exp}$ (cm$^{-1}$) & $J_j$ & $\sigma_{ij}^\mathrm{exp}$ (cm$^{-1}$) & $A_{ij}^\mathrm{exp}$ (s$^{-1}$) & removal reason \\
	\hline 
	29119  & 9 &     0	& 8	&  29119 & 	 2.32(6) &      	\\
	27851  & 8 &     0	& 8	&  27851 &	 3.26(5) &      	\\
	27556  & 7 &     0	& 8	&  27556 &	 1.75(5) &      	\\
	27014  & 9 &    0	& 8	&  27014 &	 1.19(8) &  spurious?	\\
	25012  & 8 &     0	& 8	&  25012 &	 0.37(6) &      	\\
	24906  & 7 &     0	& 8	&  24906 &	 2.93(6) &  large ratio	\\
	24229  & 9 &     0	& 8	&  24229 &	 0.92(6) &      	\\
	24204  & 8 &     0	& 8	&  24204 &	 1.76(6) &        	\\
	23832  & 8 &     0	& 8	&  23832 &	 8.80(7) &  mixed$^\mathrm{a}$ \\
	20766  & 7 &     0	& 8	&  20766 &	 0.28(5) &  small ratio	\\
	20341  & 8 &     0	& 8	&  20341 &	 1.06(5) &  small ratio	\\
	28823  & 7 &  4134	& 7	&  24688 &	 2.54(6) &      	\\
	28030  & 8 &  4134	& 7	&  23895 &	 8.80(7) &      	\\
	27984  & 7 &  4134	& 7	&  23850 &	 7.10(7) &  mixed$^\mathrm{b}$ \\
	27851  & 8 &  4134	& 7	&  23717 &	 8.10(7) &  mixed$^\mathrm{c}$ \\
	27556  & 7 &  4134	& 7	&  23422 &	 1.14(6) &      	\\
	25012  & 8 &  4134	& 7	&  20878 &	 1.08(5) &      	\\
	18022  & 8 &  4134	& 7	&  13888 &	 0.65(4) &  small ratio	\\
	15195  & 7 &  4134	& 7	&  11061 &	 0.43(5) &  small ratio	\\		
	\hline
	\multicolumn{7}{l}{$^\mathrm{a}$ mixed with level at 23878 cm$^{-1}$; $^\mathrm{b}$ with level at 27834 cm$^{-1}$; $^\mathrm{c}$ with level at 27818 cm$^{-1}$.
}
\end{longtable}

In the list of Ref.~\cite{wickliffe2000}, we can see some strong transitions whose upper level does not belong to $4f^{10}6s6p$ or $4f^{9}5d6s^2$ configurations, \textit{e.g.}~$E^{exp}=23832.060$ cm$^{-1}$, $J=8$, but is very close in energy to a $4f^{10}6s6p$ level with the same $J$, \textit{e.g.}~$E^{exp}=23877.739$ cm$^{-1}$. This suggests that the former level possess a significant $4f^{10}6s6p$ character in addition to the $4f^95d^26s$ one. However in our model, we can only describe one level, denoted $|4f^{10}6s6p\rangle\equiv|0\rangle$, which can explain the poor agreement on its Land\'e factor (see Table \ref{tab:oddlev}). Assuming that the two {}``real", mixed levels, denoted $|+\rangle$ and $|-\rangle$, are isolated from the others, we can write
\begin{eqnarray}
  |+\rangle & = & c_1|4f^{10}6s6p\rangle 
    + c_2|4f^{9}5d^26s\rangle \nonumber\\
  |-\rangle & = & -c_2|4f^{10}6s6p\rangle
    + c_1|4f^{9}5d^26s\rangle
  \label{eq:mls}
\end{eqnarray}
with $c_1^2+c_2^2=1$. We recall that the transition probability depends on the transition frequency $\omega_{ij}$ and the reduced transition dipole moment $\langle i\Vert \hat{d}\Vert j\rangle$, as $A_{ij}\propto \omega_{ij}^3| \langle i\Vert \hat{\mathbf{d}}\Vert j\rangle|^2$. In our case ($i=0$, $+$ or $-$), the transition frequencies are equal, $\omega_{0,j} \approx \omega_{+,j} \approx \omega_{-,j}$, whereas the transition dipole moments are such that $\langle 0\Vert \hat{\mathbf{d}}\Vert j\rangle|^2 = \langle +\Vert \hat{\mathbf{d}}\Vert j\rangle|^2 + \langle -\Vert \hat{\mathbf{d}}\Vert j\rangle|^2$. Therefore we compare our theoretical value $A_{0,j}^\mathrm{th}$ with the sum of experimental ones $A_{+,j}^\mathrm{exp} + A_{-,j}^\mathrm{exp}$. In Table \ref{tab:rml}, the 3 transitions labeled {}``mixed'' correspond to that situation.

Special attention should be paid to the transition between the ground level and the excited $J=9$ level at 27014.02 cm$^{-1}$. By comparison with neighboring elements like holmium, the existence of this very strong transition, in addition to the {}``usual" one with upper level $4f^{10}(^5I_8)6s6p(^1P_1^o) \, (8,1)^o_9$ at 23736.610 cm$^{-1}$, is all the more questionable, that the level at 27014.02 cm$^{-1}$ does not appear in any other transition. It is probable that this transition exists, \textit{i.e.}~its transition energy and transition probability are correct; but its lower level is probably not the ground one, and the upper level $J=9$ level at 27014.02 cm$^{-1}$ does not exist. Finally, due to strong differences between $A_{ij}^\mathrm{th}$ and $A_{ij}^\mathrm{exp}$, we excluded 5 of the last 48 transitions (one with large ratio $A_{ij}^\mathrm{th}/A_{ij}^\mathrm{exp}$, while other four with very small ratios). 

We fitted the SFs using the remaining 43 transitions, and found optimal scaling factors $f_1=0.794$, $f_2=0.923$, corresponding to a standard deviation on Einstein coefficients (see Ref.~\cite{lepers2014}, Eq.~(15)) $\sigma_A=2.66\times 10^6$ s$^{-1}$. In particular the 6 strongest transitions are calculated with a precision better than 7 \%. Then, because the experimental Einstein coefficients in Ref.~\cite{wickliffe2000} are given with uncertainties reaching to 10 \%, we made 1000 fits  in which all the $A_{ij}^\mathrm{exp}$ coefficients have a random value within their uncertainty range. We obtain optimal scaling factors with statistical uncertainties: $f_1=0.794\pm 0.006$ and $f_2=0.923\pm 0.21$. The standard deviation is therefore much more sensitive to $\langle 6s|\hat{r}|6p\rangle$ than to $\langle 4f|\hat{r}|5d\rangle$, since it involves the strongest transitions \cite{lepers2014, lepers2016}. In what follows, we take the optimal scaling factors $f_1=0.794$ and $f_2=0.923$, for which a comparison between experimental and theoretical transition probabilities involving the two lowest levels of Dy I are presented in Table \ref{tab:aik}. Using those optimal SFs, we can also calculate transition probabilities which have not been measured, and which are available upon request to the authors.

\begin{longtable} {rrrrrll}
	\caption{Comparison of Einstein-$A$ coefficients.
		The superscript {}''exp" stands for experimental values
		which are taken from \cite{wickliffe2000}. The superscript {}''th" 
		stands for the theoretical values from the present 
		calculations. The notation ($n$) stands for $\times 10^n$. Values with an asterisk ($^*$) correspond to sums of experimental Einstein coefficients (see Table \ref{tab:rml}).}
	\label{tab:aik}                           
	\endfirsthead
	\caption{Einstein-$A$ coefficients (continued)} \\
	\hline
	$E_i^\mathrm{exp}$ (cm$^{-1}$) & $J_i$ & $E_j^\mathrm{exp}$ (cm$^{-1}$) & $J_j$ & $\sigma_{ij}^\mathrm{exp}$ (cm$^{-1}$) & $A_{ij}^\mathrm{exp}$ (s$^{-1}$) & $A_{ij}^\mathrm{th}$ (s$^{-1}$) \\
	\hline
	\endhead
	\hline
	$E_i^\mathrm{exp}$ (cm$^{-1}$) & $J_i$ & $E_j^\mathrm{exp}$ (cm$^{-1}$) & $J_j$ & $\omega_{ij}^\mathrm{exp}$ (cm$^{-1}$) & $A_{ij}^\mathrm{exp}$ (s$^{-1}$) & $A_{ij}^\mathrm{th}$ (s$^{-1}$) \\
	\hline 
	25000  & 8 &     0	& 8	&  25000 & 	 1.63(5) &  2.15(5)	\\
	24709  & 7 &     0	& 8	&  24709 &	 1.92(8) &  1.91(8)	\\
	23878  & 8 &     0	& 8	&  23878 &	 2.14(8)$^*$ &  2.13(8)	\\
	23737  & 9 &     0	& 8	&  23737 &	 2.08(8) &  2.09(8)	\\
	21899  & 8 &     0	& 8	&  21899 &	 6.60(5) &  1.67(5)	\\
	21839  & 9 &     0	& 8	&  21839 &	 1.96(6) &  6.52(5)	\\
	21783  & 7 &     0	& 8	&  21783 &	 1.37(7) &  6.64(6)	\\
	21675  & 7 &     0	& 8	&  21675 &	 8.20(6) &  1.13(7)	\\
	20485  & 7 &     0	& 8	&  20485 &	 5.20(5) &  2.02(5)	\\
	19689  & 8 &     0	& 8	&  19689 &	 4.10(5) &  2.97(5)	\\
	18857  & 7 &     0	& 8	&  18857 &	 8.50(5) &  4.65(5)	\\
	18022  & 8 &     0	& 8	&  18022 &	 3.00(5) &  9.98(4)	\\
	17727  & 9 &     0	& 8	&  17727 &	 4.90(5) &  2.78(5)	\\
	17688  & 7 &     0	& 8	&  17688 &	 4.46(5) &  1.29(5)	\\
	16733  & 8 &     0	& 8	&  16733 &	 4.20(5) &  9.58(5)	\\
	16694  & 7 &     0	& 8	&  16694 &	 5.61(5) &  1.32(6)	\\
	15972  & 9 &     0	& 8	&  15972 &	 8.90(5) &  1.11(6)	\\
	15195  & 7 &     0	& 8	&  15195 &	 7.70(5) &  2.96(5)	\\
	28120  & 6 &  4134	& 7	&  23986 &	 1.92(8) &  1.83(8)	\\
	27835  & 7 &  4134	& 7	&  23701 &	 1.91(8)$^*$ &  2.03(8)	\\
	27818  & 8 &  4134	& 7	&  23684 &	 2.09(8)$^*$ &  2.06(8)	\\
	27427  & 7 &  4134	& 7	&  23293 &	 2.28(6) &  1.77(6)	\\
	25000  & 8 &  4134	& 7	&  20865 &	 1.16(6) &  4.53(5)	\\
	24907  & 7 &  4134	& 7	&  20773 &	 2.58(6) &  3.89(4)	\\
	24709  & 7 &  4134	& 7	&  20575 &	 2.59(5) &  7.50(4)	\\
	24041  & 6 &  4134	& 7	&  19906 &	 1.27(6) &  8.12(5)	\\
	21899  & 8 &  4134	& 7	&  17765 &	 1.66(5) &  1.12(5)	\\
	21783  & 7 &  4134	& 7	&  17649 &	 3.30(4) &  2.57(4)	\\
	21675  & 7 &  4134	& 7	&  17541 & 	 1.67(5) &  6.21(4)	\\
	20766  & 7 &  4134	& 7	&  16632 &	 5.90(5) &  1.10(6)	\\
	20555  & 6 &  4134	& 7	&  16421 &	 1.46(6) &  1.14(6)	\\
	20485  & 7 &  4134	& 7	&  16351 &	 5.70(4) &  3.66(4)	\\
	20341  & 8 &  4134	& 7	&  16207 &	 8.10(5) &  1.14(6)	\\
	19689  & 8 &  4134	& 7	&  15554 &	 4.50(4) &  1.97(4)	\\
	18857  & 7 &  4134	& 7	&  14723 &	 2.90(4) &  1.15(4)	\\
	17688  & 7 &  4134	& 7	&  13554 &	 1.10(5) &  1.96(4)	\\
	16733  & 8 &  4134	& 7	&  12599 &	 1.36(4) &  5.95(3)	\\
	28120  & 6 &  7051	& 6	&  21069 &	 2.59(5) &  1.53(5)	\\
	27427  & 7 &  7051	& 6	&  20376 &	 9.10(5) &  3.58(5)	\\
	24041  & 6 &  7051	& 6	&  16990 &	 6.50(4) &  2.33(4)	\\
	21783  & 7 &  7051	& 6	&  14733 &	 1.34(4) &  4.10(3)	\\
	20766  & 7 &  7051	& 6	&  13716 &	 2.90(4) &  1.43(4)	\\
	20555  & 6 &  7051	& 6	&  13504 &	 4.10(4) &  7.55(3)	\\		
	\hline	 
\end{longtable}

\section{Polarizabilities and van der Waals $C_6$ coefficients \label{sec:pc}}

The optimal set of spectroscopic data obtained in the previous section will now be used to compute polarizabilities and van der Waals $C_6$ coefficients, obtained using the sum-over-state formula inherent to second-order perturbation theory, for the ground level $4f^{10}6s^2\,^5I_8$ and the first-excited level $4f^{10}6s^2\,^5I_7$ of dysprosium. Indeed the electric-quadrupole transitions between $4f^n6s^2$ levels were suggested as candidates for optical clocks \cite{kozlov2013}, as those levels are expected to possess very similar polarizabilities.

\subsection{Polarizabilities \label{sub:rp}}

Polarizability is an important characteristic governing the optical trapping of neutral atoms, through their interaction with laser fields. The real part of the (complex) polarizability determines the depth of dipole traps or optical-lattice wells, while the imaginary part determines the photon-scattering rate, which limits the coherence and the trap lifetime for the atoms. The sum-over-state formula enables us to give the real and imaginary parts of the dynamic dipole polarizability at any trapping frequency. Because dysprosium is an open $4f$-shell atom, the trap depths and photon-scattering rate are functions of scalar, vector and tensor polarizabilities, which we give in this article.

\subsubsection{Theory of optical trapping.}

For the sake of consistency, let us recall the useful relationships of optical trapping (see \textit{e.g.}~Refs.~\cite{lepers2014, manakov1986, grimm2000}).  We assume that the atoms are in the level $|\beta JM_J\rangle$, where $J$ is the total electronic angular momentum, $M_J$ the azimuthal quantum number associated with its projection on the quantization axis $z$ and $\beta$ designates all the remaining quantum numbers. We also assume that the atoms are submitted to a laser beam of angular frequency $\omega$ and whose intensity $I(\mathbf{r})$ depends on the position $\mathbf{r}$ of the atoms. If the electric field of the laser beam is linearly polarized along the quantization axis $z$ ($\pi$ polarization), due to second-order Stark effect, it induces a potential energy $U_{M_J}^\mathrm{lin}(\mathbf{r})$ acting on the atomic center of mass,
\begin{equation}
  U_{M_J}^\mathrm{lin}(\mathbf{r}) = -\frac{1}{2\epsilon_{0}c} 
    I(\mathbf{r}) \left( \Re[\alpha_{\mathrm{scal}}(\omega)] 
    + \frac{3M_{J}^{2}-J(J+1)}{J(2J-1)}
    \Re[\alpha_{\mathrm{tens}}(\omega)] \right)
  \label{eq:u-lin}
\end{equation}
and a photon-scattering rate
\begin{equation}
  \Gamma_{M_J}^\mathrm{lin}(\mathbf{r}) = \frac{1}
    {\hbar\epsilon_{0}c} 
    I(\mathbf{r}) \left( \Im[\alpha_{\mathrm{scal}}(\omega)] 
    + \frac{3M_{J}^{2}-J(J+1)}{J(2J-1)}
    \Im[\alpha_{\mathrm{tens}}(\omega)] \right) ,
  \label{eq:gamma-lin}
\end{equation}
which both depend on the atomic Zeeman sublevel $M_J$. For a right (left) circularly polarized electric field propating along $z$ ($\sigma^\pm$ polarization), the potential and photon-scattering rate read
\begin{eqnarray}
  U_{M_J}^\mathrm{circ}(\mathbf{r}) &=& -\frac{1}{2\epsilon_{0}c} 
    I(\mathbf{r}) \left( \Re[\alpha_{\mathrm{scal}}(\omega)] 
    \pm \frac{M_{J}}{2J}\Re[\alpha_{\mathrm{vect}}(\omega)]
    \right.
  \nonumber \\
    & & \left. - \frac{3M_{J}^{2}-J(J+1)}{2J(2J-1)}
    \Re[\alpha_{\mathrm{tens}}(\omega)] \right)
  \label{eq:u-circ} \\
  \Gamma_{M_J}^\mathrm{circ}(\mathbf{r}) & = & 
    \frac{1}{\hbar\epsilon_{0}c} 
    I(\mathbf{r}) \left( \Im[\alpha_{\mathrm{scal}}(\omega)] 
    \pm \frac{M_{J}}{2J}\Im[\alpha_{\mathrm{vect}}(\omega)] 
    \right.
  \nonumber \\
    & & \left. - \frac{3M_{J}^{2}-J(J+1)}{2J(2J-1)}
    \Im[\alpha_{\mathrm{tens}}(\omega)] \right).
  \label{eq:gamma-circ}
\end{eqnarray}
In equations (\ref{eq:u-lin})--(\ref{eq:gamma-circ}), $\Re[]$ and $\Im[]$ stand for the real and imaginary parts of the scalar $\alpha_{\mathrm{scal}}$, vector $\alpha_{\mathrm{vect}}$ and tensor $\alpha_{\mathrm{tens}}$ dynamic dipole polarizabilities given by
\begin{eqnarray}
  \alpha_{\mathrm{scal}}(\omega) & = & \frac{1}{3(2J+1)} 
    \sum_{\beta''J''} \left|\left\langle\beta''J''\right\Vert 
      \mathbf{d}\left\Vert \beta J \right\rangle \right| ^{2}
  \nonumber \\
   & \times & \left( \frac{1} {E_{\beta''J''}-E_{\beta J}
    -i\frac{\hbar\gamma_{\beta''J''}}{2} - \hbar\omega}
    + \frac{1} {E_{\beta''J''}-E_{\beta J}
    -i\frac{\hbar\gamma_{\beta''J''}}{2} + \hbar\omega} \right)
  \label{eq:alpha-scal} \\
  \alpha_{\mathrm{vect}}(\omega) & = &
    \sqrt{\frac{6J}{(J+1)(2J+1)}} 
    \sum_{\beta''J''} (-1)^{J+J''} \sixj{1}{1}{1}{J}{J}{J''}
      \left|\left\langle\beta''J''\right\Vert 
      \mathbf{d}\left\Vert \beta J \right\rangle \right| ^{2}
  \nonumber \\
   & \times & \left( \frac{1} {E_{\beta''J''}-E_{\beta J}
    -i\frac{\hbar\gamma_{\beta''J''}}{2} - \hbar\omega}
    - \frac{1} {E_{\beta''J''}-E_{\beta J}
    -i\frac{\hbar\gamma_{\beta''J''}}{2} + \hbar\omega} \right)
  \label{eq:alpha-vect} \\
  \alpha_{\mathrm{tens}}(\omega) & = & 2
    \sqrt{\frac{5J(2J-1)}{(J+1)(2J+1)(2J+3)}} 
    \sum_{\beta''J''} (-1)^{J+J''} \sixj{1}{1}{2}{J}{J}{J''}
      \left|\left\langle\beta''J''\right\Vert 
      \mathbf{d}\left\Vert \beta J \right\rangle \right| ^{2}
  \nonumber \\
   & \times & \left( \frac{1} {E_{\beta''J''}-E_{\beta J}
    -i\frac{\hbar\gamma_{\beta''J''}}{2} - \hbar\omega}
    + \frac{1} {E_{\beta''J''}-E_{\beta J}
    -i\frac{\hbar\gamma_{\beta''J''}}{2} + \hbar\omega} \right)
  \label{eq:alpha-tens}
\end{eqnarray}
where $\left\langle\beta''J''\right\Vert \mathbf{d} \left\Vert \beta J \right\rangle$ is the reduced matrix element of the transition-dipole-moment operator between the level $|\beta J\rangle$ under consideration and the intermediate level $|\beta''J''\rangle$, and $\gamma_{\beta''J''}$ the radiative relaxation rate (or inverse lifetime) of the intermediate level.

\subsubsection{Results and discussion.}

To compare our results with literature, the scalar, vector and tensor static dipole polarizabilities are presented in Table \ref{tab:pola-re-im-freq}, as well as the dynamic ones for the widespread laser-trapping wavelength $\lambda=1064$ nm (corresponding to a wave number $\sigma=9398$ cm$^{-1}$). As one can notice for the ground-level scalar polarizabilities, agreement is good between the different theoretical results and which all agree well with the new measured value. The tensor static polarizability is much smaller than the scalar one in all sources; we note that our value has the same sign as the experimental one of Ref.~\cite{rinkleff1994}. As already pointed out in \cite{dzuba2011}, we observe a strong discrepancy between our 1064-nm dynamic polarizability and the experimental value of Ref.~\cite{lu2011b}. For the $^5I_7$ level there are no literature values to our knowledge. They are actually very similar to those of the ground level. For both levels, the main result obtained in our previous work on erbium \cite{lepers2014} is mostly confirmed. Regarding the real part, the vector and tensor polarizability are roughly two orders of magnitudes smaller than the scalar one, which means that the trapping potential is mostly isotropic, \textit{i.e.}~they almost do not depend on the electric-field polarization or the atomic azimuthal quantum number. By contrast, the tensor, and especially vector contributions of the imaginary part represent a significant fraction of the scalar contribution, although less significant than for erbium. This makes photon-scattering anisotropic.

\begin{table}
  \caption{Real and imaginary parts of the scalar, vector and tensor dynamic dipole polarizabilities, at $\sigma=0$ and 9398 cm$^{-1}$ ($\lambda=1064$ nm), for the ground $^5I_8$ and first excited level of dysprosium. Our results are compared with available literature values.
  \label{tab:pola-re-im-freq}}
  \centering
  \begin{tabular}{c|c|ccc|ccc}
    \hline
     \multirow{2}{*}{level} & $\sigma$  
      & \multicolumn{3}{c|}{Real part (a.u.)} 
      & \multicolumn{3}{c}{Imaginary part ($10^{-7}$~a.u.)} \\
      \cline{3-8}
      & (cm$^{-1})$ & scalar & vector & tensor & 
                      scalar & vector & tensor \\
    \hline
    $^5I_8$ & 0 & 164 & 0 & 0.835 & 30.8 & 0 & 3.40 \\
     & & 165 \cite{lide2012}, 175 \cite{chu2007} & 
     & -4.50 \cite{chu2007} & & & \\
     & & 163 \cite{dzuba2014}, 168 \cite{dzuba2016} & 
     & 1.40 \cite{rinkleff1994} & & & \\
     & 9398 & 193 & -1.49 & 1.34 
     & 49.4 & -11.3 & 5.87 \\
     &      & 116 \cite{lu2011b}, 170 \cite{dzuba2011} & & 
     & & & \\
    \hline
    $^5I_7$ & 0 & 163 & 0 & 0.723 & 30.3 & 0 & 1.66 \\
     & 9398 & 193 & -1.32 & 1.17 & 49.1 & -7.88 & 2.85 \\
		\hline
	\end{tabular}
\end{table}

\begin{figure}
	\begin{minipage} [b]{1\textwidth} \centering \includegraphics[width=9cm]{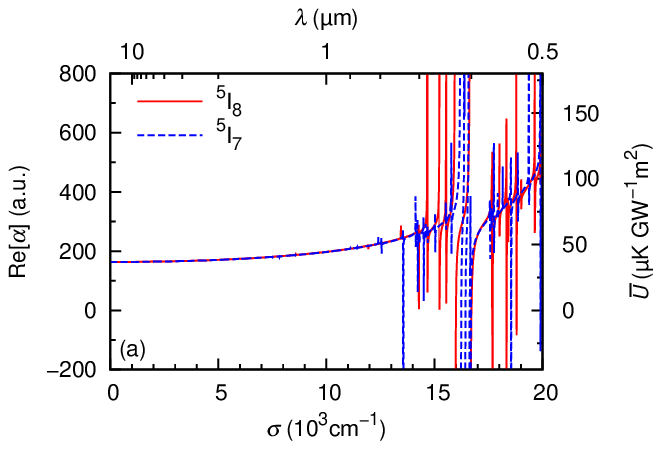} 
	\end{minipage} \\
	\begin{minipage}[b]{0.5\textwidth} \centering	\includegraphics[width=7cm]{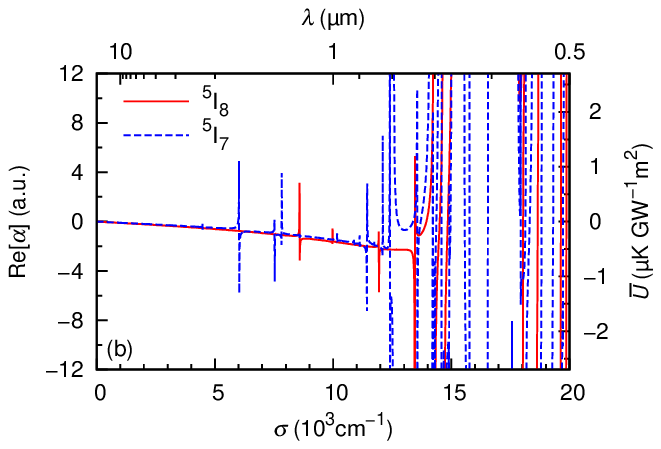}
	\end{minipage}
	\begin{minipage}[b]{0.5\textwidth} \centering	\includegraphics[width=7cm]{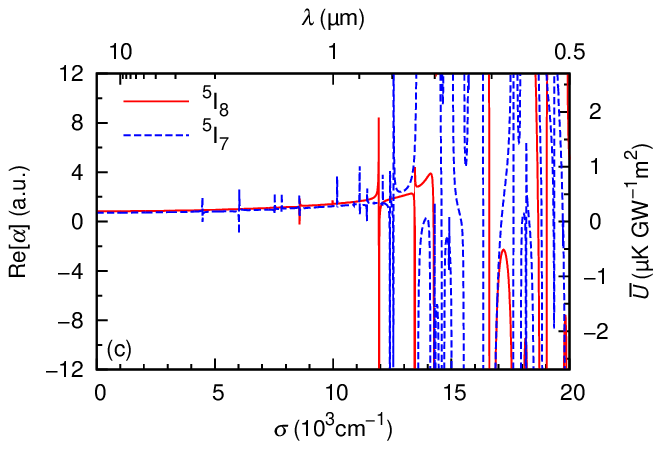}
	\end{minipage}
	\caption{(Color online) Real part of the (a) scalar, (b) vector and (c) tensor dynamic dipole polarizabilities of the $^5I_8$ and $^5I_7$ levels in atomic units and corresponding trap depths obtained for an intensity of 1 GW.m$^{-2}$, as functions of the trapping wave number $\sigma$ (or wavelength $\lambda$).
		\label{fig:pola-real-freq}}
\end{figure}

\begin{figure}
	\begin{minipage} [b]{1\textwidth} \centering \includegraphics[width=9cm]{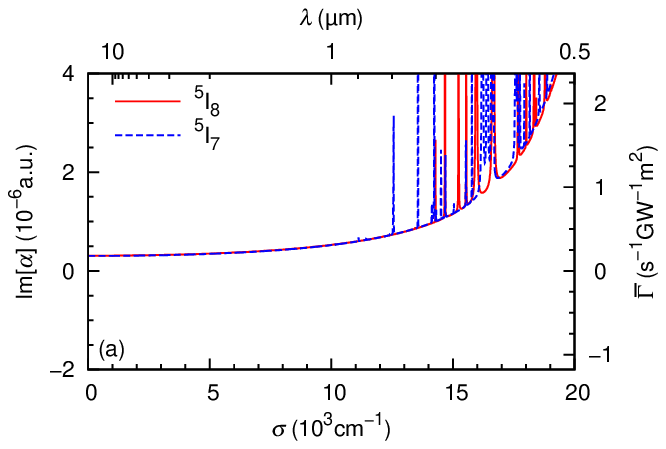} 
	\end{minipage} \\
	\begin{minipage}[b]{0.5\textwidth} \centering	\includegraphics[width=7cm]{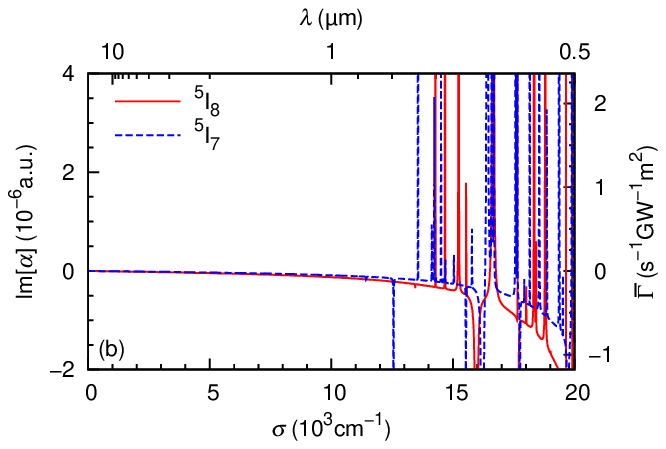}
	\end{minipage}
	\begin{minipage}[b]{0.5\textwidth} \centering	\includegraphics[width=7cm]{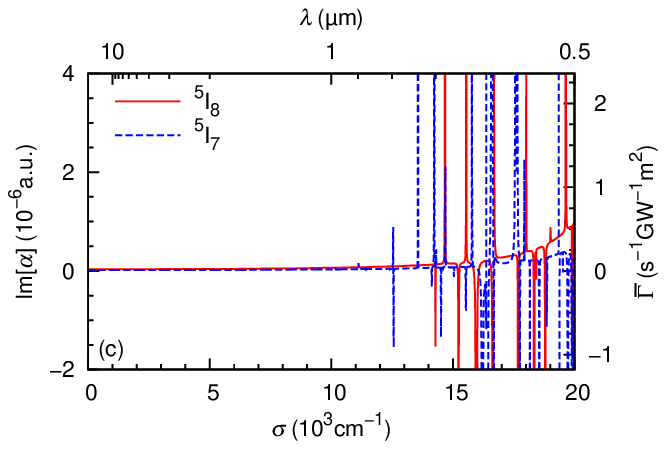}
	\end{minipage}
	\caption{(Color online) Imaginary part of the (a) scalar, (b) vector and (c) tensor dynamic dipole polarizabilities of the $^5I_8$ and $^5I_7$ levels in atomic units and corresponding trap depths obtained for an intensity of 1 GW.m$^{-2}$, as functions of the trapping wave number $\sigma$ (or wavelength $\lambda$).
		\label{fig:pola-img-freq}}
\end{figure}

Figures \ref{fig:pola-real-freq} and \ref{fig:pola-img-freq} present the real, resp.~imaginary, parts of the scalar, vector and tensor polarizabilities as functions of the field wavelength $\lambda$ and wave number $\sigma = 1/\lambda = \omega/2\pi c$ ($c$ being the speed of light). In order to facilitate experimental usage, we present our results in atomic units and also the corresponding relevant quantities in physical units. The real part of the polarizability is associated with the potential energy $\overline{U}$, in equivalent temperatures of microkelvins ($\mu$K), obtained for a laser intensity of 1~GW/m$^2$. The imaginary part is associated with the photon-scattering rate $\overline{\Gamma}$, in inverse seconds (s$^{-1}$), for the same intensity. 

Far from atomic resonances, they confirm the two phenomena described above: (i) the strong similarity between polarizabilities of the ground and first excited levels; (ii) the isotropy of the trapping potential and anistropy of photon scattering. Moreover, for wave numbers below 10000 cm$^{-1}$, the polarizabilities are essentially flat, except some very narrow peaks associated with very weak transitions. On the contrary, those background values increase (in absolute value) as the wave numbers approach the strongest transitions (see Table \ref{tab:oddlev}).

\subsection{Van der Waals $C_6$ coefficients \label{sub:vdw}}

Characterizing long-range interactions is crucial to understand the dynamics of ultracold gases. In the case of ground-level high-$Z$ lanthanide atoms, the van der Waals interaction, scaling as $R^{-6}$ ($R$ being the interatomic distance), plays a significant role, as it competes with the magnetic-dipole interaction, scaling as $R^{-3}$, for distance shorter than 100 bohr. The quadrupole-quadrupole interaction is on the contrary negligible for all distances \cite{kotochigova2011, lepers2014, frisch2015, lepers2016a}. The weak anisotropy of van der Waals interactions between lanthanide atoms is expected to be responsible for the strong density of narrow Feshbach resonances \cite{kotochigova2011, frisch2014, maier2015}.

We consider two dysprosium atoms, A and B, in the fine-structure levels $J_A$ and $J_B$ of the lowest multiplet $\beta_A = \beta_B = \,^5I$ of the ground configuration $4f^{10}6s^2$. The Zeeman sublevels are characterized by the azimuthal quantum numbers $M_{J_A}$ and $M_{J_B}$, taken with respect to the quantization axis $z$ of the spaced-fixed frame. In this frame, the two atoms can perform an end-over-end rotation, characterized by the partial-wave quantum numbers $L$ and $M_L$. It can be shown \cite{stone1996, kaplan2006} that the van der Waals interaction is represented by an effective operator $\hat{\mathrm{W}}(\mathbf{R})$, depending the vector $\mathbf{R}$ joining the two atoms (from A to B). Expressed in the basis $\{|\beta_A J_A M_{J_A} \beta_B J_B M_{J_B} LM_L\rangle\}$, the matrix elements of $\hat{\mathrm{W}}(\mathbf{R})$ can be expressed as functions of a few parameters $C_{6,k_A k_B}$ and of angular factors,
\begin{eqnarray}
  & & \langle \beta_A J_A M'_{J_A} \beta_B J_B M'_{J_B} L'M'_L|
    \hat{\mathrm{W}}(\mathbf{R})
    |\beta_A J_A M_{J_A} \beta_B J_B M_{J_B} LM_L\rangle
  \nonumber \\
  & = & \frac{C_{6,00}}{R^6} \delta_{M_{J_A} M'_{J_A}}
    \delta_{M_{J_B} M'_{J_B}} \delta_{LL'} \delta_{M_L M'_L}
    + 15 \sum_{k_A,k_B=0}^2 \left(2k_A+1\right) 
    \left(2k_B+1\right)
  \nonumber \\
   & \times & \frac{C_{6,k_Ak_B}}{R^6}
    \sqrt{\frac{2L_A+1}{2L'_A+1}}
    \sum_{k=|k_A-k_B|}^{k_A+k_B} C_{L0k0}^{L'0} C_{2020}^{k0}
    \ninej{1}{1}{2}{1}{1}{2}{k_A}{k_B}{k}
  \nonumber \\
   & \times & 
    \sum_{q_A=-k_A}^{k_A} \sum_{q_B=-k_B}^{k_B}  
    \sum_{q=-k}^{k} \left(-1\right)^q 
    C_{k_Aq_Ak_Bq_B}^{kq} C_{LM_L,k,-q}^{L'M'_L}
    C_{J_AM_{J_A}k_Aq_A}^{J_AM'_{J_A}}
    C_{J_BM_{J_B}k_Bq_B}^{J_BM'_{J_B}}
  \label{eq:w}
\end{eqnarray}
where $C_{a\alpha b\beta}^{c\gamma}$ is a Clebsch-Gordan coefficient and the number between curly brackets is a Wigner 9j-symbol \cite{varshalovich1988}. The coefficients $C_{6,k_Ak_B}$ read
\begin{eqnarray}
  C_{6,k_Ak_B} & = & -\frac{2}{\sqrt{(2J_A+1)(2J_B+1)}}
    \sum_{\beta''_A J''_A \beta''_B J''_B}
    \left(-1\right)^{J_A+J''_A+J_B+J''_B}
    \sixj{1}{1}{k_A}{J_A}{J_A}{J''_A}
  \nonumber \\
    & & \times \sixj{1}{1}{k_B}{J_B}{J_B}{J''_B}
    \frac{\left|\left\langle\beta''_AJ''_A\right\Vert 
      \mathbf{d}_A\left\Vert \beta_A J_A \right\rangle
      \left\langle\beta''_BJ''_B\right\Vert \mathbf{d}_B 
      \left\Vert \beta_B J_B \right\rangle\right|^{2}}
    {E_{\beta''_AJ''_A}-E_{\beta_AJ_A} 
    +E_{\beta''_BJ''_B}-E_{\beta_BJ_B}}
  \label{eq:c6}
\end{eqnarray}
where $\{:::\}$ is a Wigner 6j-symbol \cite{varshalovich1988}. In equation (\ref{eq:w}), the indexes $k_A$ and $k8B$ are such that $k_A+k_B$ is non-zero and even. Giving diagonal matrix elements, the coefficient $C_{6,00}$ is referred to as isotropic, while the other $C_{6,k_Ak_B}$ are called anisotropic.

\begin{table}
  \centering
  \begin{tabular}{c|ccc}
    \hline
     $k_A$, $k_B$ & $^5I_8$ -- $^5I_8$ & $^5I_8$ -- $^5I_7$
      & $^5I_7$ -- $^5I_7$ \\
    \hline
     0, 0 & -2.27(3)  & -2.27(3)  & -2.27(3)  \\
     1, 1 & -8.13(-2) & -7.84(-2) & -7.56(-2) \\
     2, 0 &  1.36(0)  &  1.36(0)  &  1.05(0)  \\
     0, 2 &  1.36(0)  &  1.05(0)  &  1.05(0)  \\
     2, 2 & -5.23(-3) & -4.74(-3) & -4.32(-3) \\
    \hline
  \end{tabular}
  \caption{$C_6$ coefficients in atomic units, characterizing the van der Waals interactions between dysprosium atoms in the ground $^5I_8$ or first excited level $^5I_7$, as functions of the pairs of indexes $k_A$, $k_B$ (see text). The case $k_A=k_B=0$ corresponds to the so-called isotropic $C_6$ coefficient \cite{stone1996}.
  \label{tab:c6}}
\end{table}

Table \ref{tab:c6} displays our calculated $C_{6,k_Ak_B}$ coefficients. Similarly to the case of erbium \cite{lepers2014}, the isotropic one strongly dominates the anisotropic ones. Furthermore, after diagonalization of equation (\ref{eq:w}), one obtains $C_6$ coefficients ranging from -2277 to -2271 a.u., hence roughly 20 \% larger than those of Ref.~\cite{kotochigova2011}.

\section{Concluding remarks \label{sec:conclu}}

In this article, we have characterized the optical trapping of ultracold dysprosium atoms, by calculating the real and imaginary parts of the scalar, vector and tensor contributions to the dynamic dipole polarizabilities of the ground level and the first excited level. The results indicate that the trapping potential, associated with the real part, is essentially isotropic, while the photon-scattering rate, associated with the imaginary part,  exhibits a noticeable anisotropic behavior. These conclusions for the ground level are similar to our previous work on erbium, even though the anistropy of photon scattering is clearly less pronounced. The reasons for this difference are still under examination, and we expect our future work on neighboring elements (holmium and thulium) to clarify those reasons. For the 1064-nm real part of the polarizability, our results support the previous theoretical value rather than the experimental one. We also find that the polarizabilities of the ground and the first excited levels are very close, which makes them interesting candidates for optical clocks.

In order to calculate polarizabilities, we have modeled the spectrum -- energy levels and transition probabilities -- of neutral dysprosium. We have performed a systematic adjustment of theoretical and experimental Einstein-$A$ coefficients. In comparison with our previous work on neutral erbium, we have incorporated in the fit transitions toward the first excited level of dysprosium, which gives us more reliable values of the mono-electronic transition dipole moments. In addition, our detailed spectroscopic study allows us for putting into question the existence of the tabulated level $J=9$ at 27014.02 cm$^{-1}$, which is however expected to have a strong transition probability towards the ground level.

\section*{Acknowledgements}

The authors acknowledge support from {}``IFRAF/Dim Nano-K'' under the project InterDy, and from {}``Agence Nationale de la Recherche'' (ANR) under the project COPOMOL (contract ANR-13-IS04-0004-01). SN acknowledges support from European Union (ERC UQUAM) and PSL Research University (MAFAG project).

\appendix

\section{Energy parameters}

This appendix presents the optimal parameters after the least-sq		uare fitting procedure of experimental and theoretical energies. Table \ref{tab:parev} deals with even-parity levels, and Table \ref{tab:parod} with odd-parity levels. The effective parameters $\alpha$, $\beta$, $\gamma$ with fixed values are taken for our previous work \cite{lepers2013}.

\begin{longtable}{crrrrr}
	\caption{\label{tab:parev} Fitted one-configuration parameters 
		(in cm$^{-1}$) for even-parity configuration of Dy I compared 
		with HFR radial integrals. The scaling factors are 
		$SF(P)=P_\mathrm{fit}/P_\mathrm{HFR}$, except for $E_\mathrm{av}$ 
		when they are $P_{fit}-P_{HFR}$. Some parameters are constrained 
		to vary in a constant ratio $r_n$, indicated in the second column 
		except if {}`fix' appears in the second or in the {}`Unc.' columns. In this case, 
		the parameter $P$ is not adjusted.
		The {}`Unc.' columns named after {}`uncertainty' present the standard error on each parameter after the fitting procedure.}
	\\
	\hline
	& &   \multicolumn{4}{c}{4f$^{10}6s^2$}  \\
	\cline{3-6} 
	\endfirsthead
	\caption{Fitted parameters in Dy I (continued)} \\
	\hline
	& & \multicolumn{4}{c}{Fitted parameters}  \\
	\cline{3-6} 
	Param. $P$      &  Cons.   & $P_\mathrm{fit}$ & Unc. & $P_\mathrm{HFR}$   & $SF$  \\
	\hline
	\endhead                                                                  
	Param. $P$      &  Cons.   & $P_\mathrm{fit}$ & Unc. & $P_\mathrm{HFR}$ &  $SF$   \\
	\hline
	$E_\mathrm{av}$ &          &           41851  &  68  &            0     & 41851   \\
	$F^2(4f4f)$     &  $r_{1}$ &           83325  & 648  &          115093  & 0.724  \\
	$F^4(4f4f)$     &  $r_{2}$ &           56562  & 1128 &           71831  & 0.787  \\
	$F^6(4f4f)$     &  $r_{3}$ &           46457  & 689  &           51572  & 0.901   \\
	$\alpha$        &      fix &            20.0  &      &                  &           \\
	$\beta$         &      fix &            -650  &      &                  &      \\
	$\gamma$        &      fix &            2000  &      &                  &           \\
	$\zeta_{4f}$    &  $r_{4}$ &            1770  &   2  &           1845   & 0.959  \\ 
	\hline	
\end{longtable}

\begin{longtable}{crrrrrrrrr}
	
	\caption{\label{tab:parod} Same as Table \ref{tab:parev} but for odd-parity levels.}
	\\
	\hline
	& &   \multicolumn{4}{c}{4f$^{9}5d6s^2$} & \multicolumn {4}{c}{4f$^{10}6s6p$} \\
	\cline{3-6} \cline{7-10}
	\endfirsthead
	\caption{Fitted parameters in Dy I (continued)} \\
	\hline
	& & \multicolumn{4}{c}{Fitted parameters} & \multicolumn{4}{c}{Fitted parameters} \\
	\cline{3-6} \cline{7-10}
	Param. $P$      &  Cons.   & $P_\mathrm{fit}$ & Unc. & $P_\mathrm{HFR}$   & $SF$ &  $P_\mathrm{fit}$  & Unc.   & $P_\mathrm{HFR}$   & $SF$ \\
	\hline
	\endhead                                                                  
	Param. $P$      &  Cons.   & $P_\mathrm{fit}$ & Unc. & $P_\mathrm{HFR}$ &  $SF$  &  $P_\mathrm{fit}$ & Unc.  & $P_\mathrm{HFR}$ &  $SF$ \\
	\hline
	$E_\mathrm{av}$ &          &           68280  &  97  &            5400  & 62880  &            61716 &   66  &            14794 & 46922 \\
	$F^2(4f4f)$     &  $r_{1}$ &           91462  & 471  &          122573  & 0.746  &            85931 &  443  &           115161 & 0.746 \\
	$F^4(4f4f)$     &  $r_{2}$ &           61462  & 739  &           76869  & 0.800  &            57486 &  691  &            71875 & 0.800 \\
	$F^6(4f4f)$     &  $r_{3}$ &           48518  & 462  &           55292  & 0.877  &            45282 &  431  &            51604 & 0.877 \\
	$\alpha$        &      fix &            20.0  &      &                  &        &             20.0 &       &                  &       \\
	$\beta$         &      fix &            -650  &      &                  &        &             -650 &       &                  &       \\
	$\gamma$        &      fix &            2000  &      &                  &        &             2000 &       &                  &       \\
	$\zeta_{4f}$    &  $r_{4}$ &            1901  &   3  &           1975   & 0.963  &             1777 &    3  &             1846 & 0.963 \\ 
	$\zeta_{5d}$    & $r_{10}$ &             727  &  12  &            890   & 0.817  &                  &       &               &      \\   
	$\zeta_{6p}$    &  $r_{6}$ &                  &      &                  &        &             1372 &   12  &              947 & 1.449 \\ 
	$F^1(4f5d)$     & $r_{11}$ &             658  &  99  &                  &        &                  &       &                  &       \\ 
	$F^2(4f5d)$     & $r_{12}$ &           15708  & 132  &           20992  & 0.748  &                  &       &                  &       \\ 
	$F^4(4f5d)$     & $r_{13}$ &           11704  & 223  &            9652  & 1.213  &                  &       &                  &      \\
	$F^1(4f6p)$     &  $r_{7}$ &                  &      &                  &        &              112 &   38  &                  &     \\   
	$F^2(4f6p)$     &  $r_{8}$ &                  &      &                  &        &             3093 &  268  &             3386 & 0.913  \\
	$G^1(4f5d)$     & $r_{14}$ &            5785  & 110  &            9181  & 0.630  &                  &       &                 & \\        
	$G^2(4f5d)$     & $r_{15}$ &            2071  & 255  &                  &        &                  &       &                 &  \\       
	$G^3(4f5d)$     & $r_{16}$ &            6731  & 288  &            7278  & 0.925  &                  &       &                 &  \\
	$G^4(4f5d)$     & $r_{17}$ &            4003  & 393  &                  &        &                  &       &                 & \\        
	$G^5(4f5d)$     & $r_{18}$ &            5480  & 236  &            5501  & 0.996  &                  &       &                 &      \\
	$G^3(4f6s)$     &  $r_{9}$ &                  &      &                  &        &             1132 &   51  &             1688 & 0.670 \\
	$G^2(4f6p)$     &  $r_{5}$ &                  &      &                  &        &             1032 &   17  &              774 & 1.333 \\        
	$G^4(4f6p)$     &  $r_{5}$ &                  &      &                  &        &              900 &   15  &             675 & 1.333  \\ 
	$G^1(6s6p)$     &      fix &                  &      &                  &        &            10292 &       &           23189 & 0.444 \\
	configuration-interaction & & \multicolumn{4}{c}{$4f^{9}5d6s^2-4f^{10}6s6p$} \\
	\hline
	\cline{3-6}
	$R^{1}(5d6s,4f6p)$ & $r_{5}$ &  -3545 &   58 &  -4648 & 0.763  \\
	$R^{3}(5d6s,6p4f)$ & $r_{5}$ &   -748 &   12 &   -980 & 0.763  \\                                                        
	\hline	
\end{longtable}

\section*{Bibliography}


\end{document}